\documentclass[12pt]{article}
\usepackage{graphicx}
\usepackage{layout}

\newcommand{\dphi}{\mbox{$\Delta\phi$}}
\newcommand{\pt} {\mbox{$p_T$}}
\newcommand{\kt} {\mbox{$k_T$}}
\newcommand{\piz} {\mbox{$\pi^{0}$}}
\newcommand{\gam} {\mbox{$\gamma$}}

\newcommand{\raa} {\mbox{$R_{AA}$}}
\def\sqrts{$\sqrt{s}$ }
\def\sqrtsNN{$\sqrt{s_{NN}}$ }

\def\raa{$R_{AA}$}
\def\auau{$Au+Au$}
\def\pp{$p+p$}
\def\rg{$R_{\gamma}$}
\def\ptgpth{$p_{T,\gamma},p_{T,h}$}

\def\pout{$p_{out}$}
\def\xhat{$\hat{x}$}
\def\iaa{$I_{AA}$}

\begin{document}
{\small \it Hard Probes 2008 Conference Proceedings.  June 9th, 2008. Illa da Toxa, Spain}
\vspace{12pt}



\centerline{\Large \bf Two-particle Direct Photon-Jet Correlation} \centerline{\Large \bf Measurements in PHENIX}

\vspace{12pt} 


\centerline{{\bf J.~Frantz}$^{\rm a}$ for the PHENIX Collaboration}

\vspace{12pt} 


\centerline{$^{\rm a}$ State University of New York (SUNY), Stony Brook, Stony Brook, NY, U.S.A }

\vspace{8pt} 

\vspace{12pt} 


\centerline{Contact e-mail: {\it jfrantz@skipper.physics.sunysb.edu}}

\title{Two-particle Direct Photon-Jet Correlation Measurements in PHENIX}
\author{Justin Frantz 
for the PHENIX Collaboration}


%
\date{Received: date / Revised version: date}
%
\abstract{
Various 2-particle direct photon-hadron correlation yields in $p+p$ and $Au+Au$ collisions at \sqrtsNN= 200 GeV are presented.  The per-trigger yield of direct photon hadron pairs from direct-photon-jet correlations is obtained by a statistical subtraction of the decay photon pairs from inclusive photon-hadron sample.  The decay photon per-trigger yields are estimated from the measured $\pi^0$-hadron by means of a Monte Carlo based calculation which takes into account decay kinematics and detector response.  Under the assumption that the suppression is nearly \pt\ independent using a specific averaging scheme, we find a ratio of \auau\ to \pp\ per-trigger photon yields $I_{AA}$ averaged over all available \pt\ bins  consistent with the single particle suppression level \raa, which can be interpreted as a qualitative confirmation of the basic geometrical picture of jet suppression at RHIC.  The application of the event by event photon isolation cuts in $p+p$ results our highest precision measurement yet, and allows for precision studies of the baseline fragmentation function $D(z)$ and also a determination of the apparent intrinsic $k_T$, or non-zero transverse momentum of the original collision partons.  With a model dependent extraction method, the average $\sqrt{<k_T^2>}$ at this \sqrts\ in \pp\ is found to be approximately 2.5-3 GeV, consistent with analysis of di-hadron (di-jet) correlations \cite{ppg029}.   Finally, we present a unique direct measurement of prompt photons from jet fragmentation.

} 
%
%
\section{Introduction}
\label{intro}

At the Relativistic Heavy Ion  Collider (RHIC), experimental results from the have established the formation of hot and dense matter of a fundamentally new nature in Au+Au collisions at $\sqrt{s_{NN}}$=200 GeV~\cite{whitepaper}.  One of the most important probes of this dense colored matter is energy loss by hard partons leading to a suppression of normal jet production of hard (E $> \sim 1-2$ GeV) particles.  As a complement to di-hadron correlations, which can directly access di-jet production and their structure \cite{stardijet}, \cite{phenix_ppg083}, direct photon-hadron correlations can be used to study photon-jet production in the medium without various biases and complications.  This is due to the fact that since it lacks color charge, the photon escapes the dense QCD matter without interacting.  For this reason, using high $p_T$ prompt photons, which have been demonstrated to be unmodified when traversing the medium due to their lack of color charge \cite{ppg042},  as triggers in $\gamma$ - hadron ($i.e.$ $\gamma$-jet) correlation studies has long been considered to be a "golden channel" to study energy loss.  When the photon trigger is a product of the dominant Leading Order QCD Compton scattering process, the photon's 4-momentum should be the same as the opposite quark parton's. Therefore the photon in its primordial state can be used to directly measure the fragmentation functions of the opposite jet which itself is modified by final state effects in the case of Au+Au collisions.

The above picture is based on a Leading Order (LO) approximation and interpretation of events where there are always only two "clean" outgoing products, di-parton (\emph{g, q}) or photon-parton for \gam-jet, which are perfectly balanced in momentum in the direction transverse to the incoming colliding particles.   In reality there are a number of complications that could make this interpretation incorrect that need to be understood even in the baseline elementary \pp\ collisions.  For example, Next to Leading Order (NLO)/fragmentation contributions of hard prompt photons in di-jet events obviously have different properties, in particular in $Au+Au$ are expected to follow similar suppression effects as the di-jet events which makes making quantifying the baseline probability for their appearance in the overall \gam-jet event samples crucial.  Another example is the role of an apparent intrinsic net transverse momentum $k_T$ of the incoming colliding parton-parton system potentially due to initial state effects or non-perturbative or higher order gluon radiation, which at the value discovered in dihadron correlations \cite{ppg029} can considerably change the actual momentum transfer ($Q^2$) being sampled in the kinematic ranges currently being studied.   Thus, before studying \gam-jet production A+A, one needs to understand these intrinsic complications to the LO picture in $p+p$ to gauge their effect on calculations and interpretations of energy loss observables.

This analysis was performed using approximately 950 million Au+Au minimum bias events from the Run 4 data set and 471 million level-1 photon-triggered events from the p+p Run 5 and Run 6 data sets.  The Beam-Beam Counters (BBC) and Zero-Degree Calorimeters (ZDC) are used to trigger the minimum bias data.  These detectors are also used to estimate the collision centrality.  The p+p photon trigger requires that a module in the Electromagnetic Calorimeter (EMC) fire in coincidence with the BBC.

The PHENIX central arms, each covering the 0.7 units of pseudo-rapidity around mid-rapidity and $90^\circ$ in azimuth, contain charge particle tracking chambers and electromagnetic calorimetry.  The EMC consists of two types of detectors,  six sectors of lead-scintillator sampling calorimeters and two of lead-glass Cherenkov calorimeters.  The fine segmentation of the EMC ($\sim 0.01 \times 0.01$ for PbSc and $\sim 0.008 \times 0.008$ for PbGl) allows for the reconstruction of the $2\gamma\ $ \piz\ and $\eta$ decays to \pt\ $> 20$ GeV.

Charged hadrons are detected using a drift chamber at a radial distance of 2.0 m and and a multi-wire proportional chambers (PC1) at a distance of 2.5 m.  The momentum resolution was determined to be $0.7\%  \bigotimes 1.0 \%p$ (GeV/c).  Secondary tracks from decays and conversions are suppressed by matching tracks to hits in a second multi-wire chamber (PC3) and the EMC, both at distance of $\sim 5.0$ m.   Track projections to the EMC plane are used to veto charged hadrons which shower in the EMC.

Two-particle correlations are performed by measuring the azimuthal angle between photon triggers and charged hadrons.  The correlation function

$$C(\Delta\phi) \equiv \frac{N_{pairs}(\Delta\phi)}{ N_{mixed}(\Delta\phi)}$$

corrects for the limited acceptance of photon-hadrons pairs by dividing out the mixed events distribution.  The correlation function is decomposed via the two-source model where the jet correlation is superimposed on an underlying event which is modulated by elliptic flow.  Hence, the jet function is expressed as $J(\Delta\phi) \equiv C(\Delta\phi) - b_0(1+2\langle v_2^{\gamma} v_2^h \rangle \cos{2\Delta\phi})$.  The underlying event level, $b_0$, is determined by the ZYAM procedure, described elsewhere \cite{ppg032}.

We define a direct photon as any photon not from a decay process.  It follows that the per-trigger yield ($Y \equiv 1/N_{trigger}\ J(\Delta\phi)$) for direct photons may be obtained by a statistical subtraction of the decay per-trigger yield from the inclusive per-trigger yield according to:

\begin{equation}
  Y_{direct} = \frac{R_{\gamma}  Y_{inclusive} - Y_{decay}}{R_{\gamma} -1}
  \label{eq:subform}
\end{equation}

where \rg $\equiv N_{inclusive}/N_{decay}$.  \rg\ is determined for Au+Au collisions in \cite{ppg042} and is derived from the \piz\  \cite{ppg063} and direct photons spectra in p+p \cite{ppg060}.  \ The direct photon yields from the statistical subtraction method do not, by definition, exclude photons from jet fragmentation or medium induced photon production.

The decay photon per-trigger yields are determined from the parent (\piz\ or $\eta$) per-trigger yields via a Monte Carlo procedure.  A flat distribution of parent mesons are decayed into the PHENIX aperture uniformly in the z and phi directions.  This determines the mapping of the parent to daughter \pt, $\wp(p_T^{\pi^{0}} \rightarrow p_T^{\gamma})$, where $p_T^{\gamma}$ is smeared by the detector resolution.   In order to reproduce the correct \pt dependence of the decay photon distribution W is applied as a weight factor to the parent meson-hadron \pt distribution on a pair-by-pair basis.  The finite reconstruction efficiency of the parent mesons is corrected for using the published PHENIX \pt\ spectra \cite{ppg063}, \cite{ppg054}.  The decay per-trigger yield from \piz's can then be expressed in terms of the parent per-trigger yield.

\section{Constraining Direct Photon-Jet Production in Elementary \pp\ Collisions}
\label{sec:constrain_pp}

\subsection{Isolation Cut Analysis}
\label{sec:cons_pp_iso}

In addition to the above statistical subtraction method, standard photon isolation cuts were applied event by event in a new analysis  in order to dramatically reduce the contamination of di-jet events with \piz\ decay or fragmentation photons in the \gam-hadron event sample.  To be considered isolated, the sum of \pt from all tracks and and EMCal energies must be  $<  0.1E_\gamma$ in a cone around the photon of size $\Delta R = \sqrt{\Delta\phi^2+\Delta\eta^2} = 0.5$. Statistical subtraction of the remaining contribution for isolated \piz\ production is achieved through a modified version of the statistical method above where isolated \piz\ are used as input to the decay-photon calculation.

\begin{figure}
\centering
\includegraphics[width=0.5\textwidth]{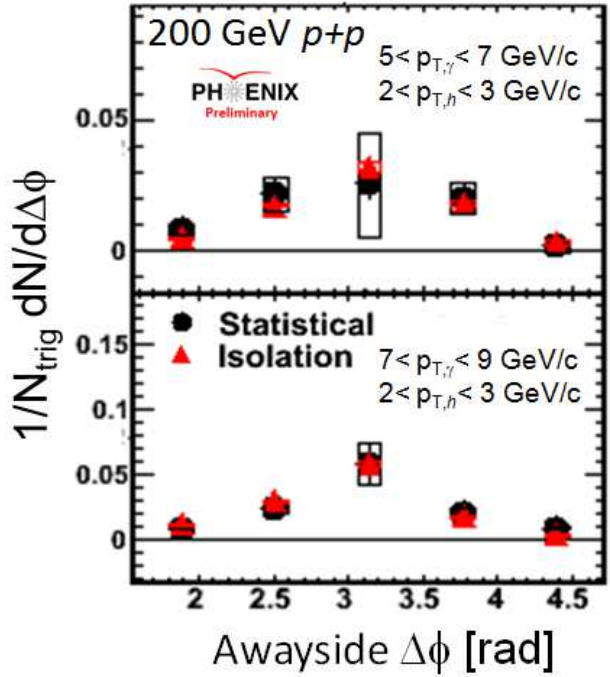}
\caption{For two trigger photon bins as indicated on the figure, the awayside per-trigger yield vs. \dphi\ for the statistical and isolation cut analysis.  The two methods give consistent results with the isolation method having much improved precision due to having only a small decay photon background subtraction.}
\label{fig:isodemo}
\end{figure}

As in the statistical method the analysis is performed as a function of \dphi.  We find good agreement with the statistical method results as shown in figure \ref{fig:isodemo}.  Since the statistical subtraction method includes, in principle, the contribution from fragmentation photon triggers, the agreement between the two methods places a constraint on the magnitude of such a contribution.

\subsection{Per trigger yields and Awayside Jet Fragmentation Function Analysis}
\label{sec:cons_pp_results}

\begin{figure}
\includegraphics[width=\textwidth]{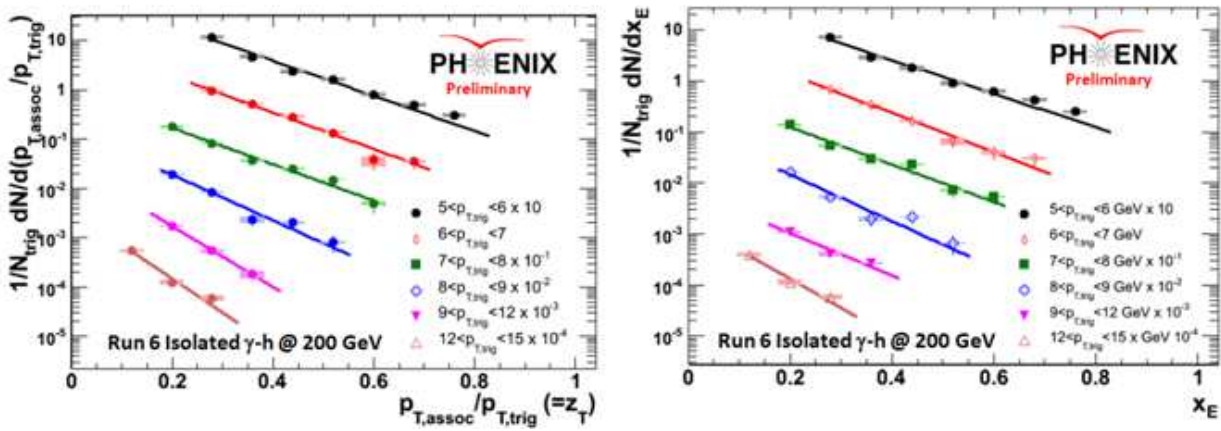}
\caption{Per trigger yields of direct photon-hadron pairs as a function of $x_{E}/z_{T}$ for various selections of trigger \pt.  The data have been scaled by factors of 10 for visibility.}
\label{fig:zt_xe}       
\end{figure}

\begin{figure}
\includegraphics[width=0.5\textwidth]{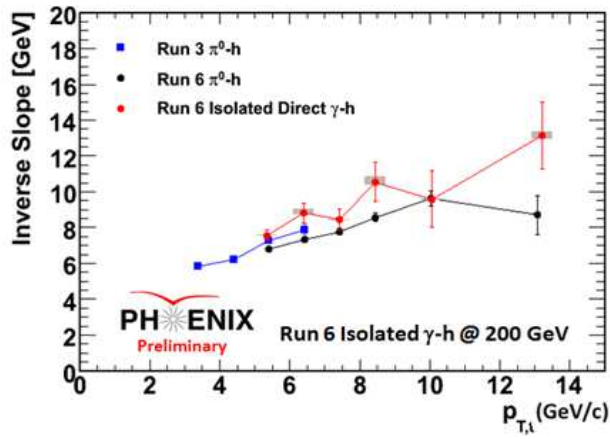}
\centering
\caption{The inverse slope parameter of exponential fits to the $x_{E}$ distributions for $\pi^{0}$ direct photon triggers.}
\label{fig:xeslopes}
\end{figure}

Results from the isolation cut analysis are shown in figure  \ref{fig:zt_xe}, plotted in terms of the fragmentation variables $z_T = p_T^{assoc}/p_T^{\gamma} $ and $x_E = z_T\cos{\Delta\phi}$ (taking only the component of the associated hadron momentum that is in the same direction as the trigger photon).  We measure in 6 photon trigger \pt\ bins covering the range 5-15 GeV/c.  The inverse slope parameter of exponential fits to the $x_{E}$ distribution with $\pi^{0}$ and direct photon triggers are shown in figure \ref{fig:xeslopes}.  Comparisons with theoretical calculations are currently underway.

\subsection{Model Dependent Determination of $k_T$}
\label{sec:cons_pp_kt}

\begin{figure}[htp]
\centering
\includegraphics[width=0.5\textwidth]{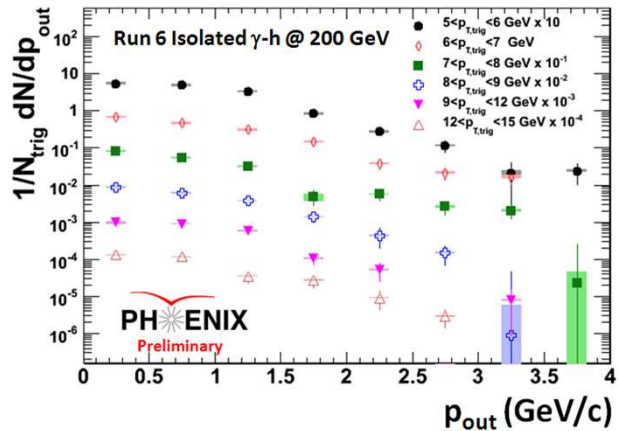}
\caption{\pout\ distributions for various values of trigger \pt.  The data have been scaled by factors of 10 for visibility.}
\label{fig:pout}
\end{figure}

Also in \cite{ppg029}, a method for extracting the apparent intrinsic \kt\ itself was described and used to extract a value of \kt = $ 2.68\pm0.35$ for \piz-h correlations.  Using isolated direct photon triggers, the method is simplified and can be also be used to extract a \kt\ value in this channel.  Please refer \cite{ppg029} for details of the method.  The simplification occurs due to the lack of an additional fragmentation variable on the nearside, i.e $p_{Tt} \equiv \hat{p}_{Tt} \equiv p_{T\gamma}$.   The measured experimental variable \pout\, whose distribution for various trigger \pt\ bins is shown in figure \ref{fig:pout}, is defined as $p_{Ta} \sin{\Delta\phi}$, or the associated hadron's transverse momentum component perpendicular to the trigger photon direction, and is proportional to the \kt\ on an event by event basis.

\begin{figure}
\includegraphics[width=\textwidth]{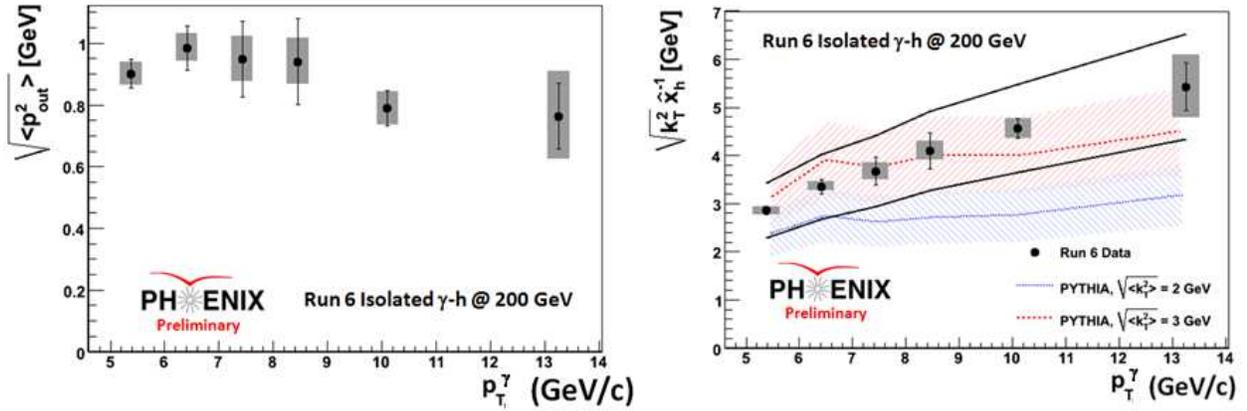}
\caption{a) $\sqrt{<p_{out}^{2}>}$ vs trigger \pt\ for direct photons b) \kt/\xhat\ vs trigger \pt\ for direct photons along with calculations from PYTHIA $6.3$.}
\label{fig:pout_kt}       
\end{figure}

By finding the average \pout\ for each trigger photon \pt\ bin as shown in \ref{fig:pout_kt} $a)$ one can extract the quantity average \kt/\xhat\ where \xhat\ is just the ratio of the true awayside \textit{jet} momentum to that of the direct photon trigger.  This can be extracted from other measurements as in \cite{ppg029} and eliminated, leaving a pure measurement of the \kt.  For now, we simply use the event generator PYTHIA $6.3$ to generate direct photon processes with initial and final state radiation turned off, and phenomenological \kt\ parameters as shown on \ref{fig:pout_kt} $b)$ to extract \xhat\ and make a comparison to PYTHIA, which shows that for PYTHIA-like distributions of \xhat\, a \kt\ parameter in the vicinity of 2.5-3 GeV gives values similar to the data, well-consistent with the value of 2.68 found for di-hadron di-jet correlations.

This implies that the same \kt\ bias effects in the di-hadron correlations exist also in the direct photon-jet correlations.  This makes the comparison with the perturbative QCD calculations, which at only NLO should \emph{not} contain such \kt\ bias modifications even more important, as well as understanding in energy loss models of \auau\ how such a large value of imbalance between the trigger photon and awayside jet might alter energy loss interpretations.

\subsection{Direct Measurement of Prompt Photons from Jet Fragmentation}
\label{sec:cons_pp_frag}

\begin{figure}[tb]
\centering
\includegraphics[width=0.5\textwidth]{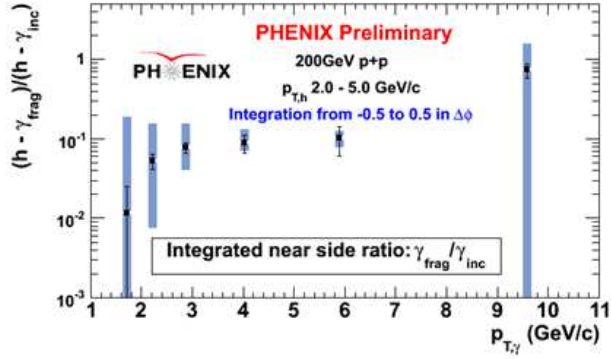}
\caption{Fraction of hadron-photon that contain a prompt (fragmentation) photon }
\label{fig:alifrag}
\end{figure}

PHENIX has made further strides towards understanding complications to the LO intepretations of direct photon-jet correlations, in studying the contribution of single prompt photons that occur in di-jet events, the so-called fragmentation photons produced by NLO hard photon radiation or non-perturbative fragmentation processes.  A direct measurement technique discussed in \cite{ali} has been used to measure the angular distribution of such photons with respect to trigger \emph{hadrons} in events where high \pt\ photons are tagged to only be nearby such hard hadrons--an \emph{anti}-isolation selection.  Integrating this distribution one can find the fraction of such photons to total hadron-correlated photons from all sources including decay, shown in figure \ref{fig:alifrag} which can possibly be related to the fraction of the total direct photon production rate due to these fragmentation photons.  However due to the restricted kinematic region of the measurement, interpretations of this fraction and the angular distribution itself need input from sophisticated higher order pQCD calculations (which likely do not yet exist) and thus close attention from the theory community.  Nonetheless, the measurement is exciting because it provides the first step towards making the measurement in the \auau\ environment where several interesting predictions of jet-medium enhancement of the rate of production of such photons exist.

\section{Constraining Energy Loss theory in Au+Au with $\gamma$-jet}
\label{sec:cons_auau}

\begin{figure}[tb]
\centering
\includegraphics[width=0.5\textwidth]{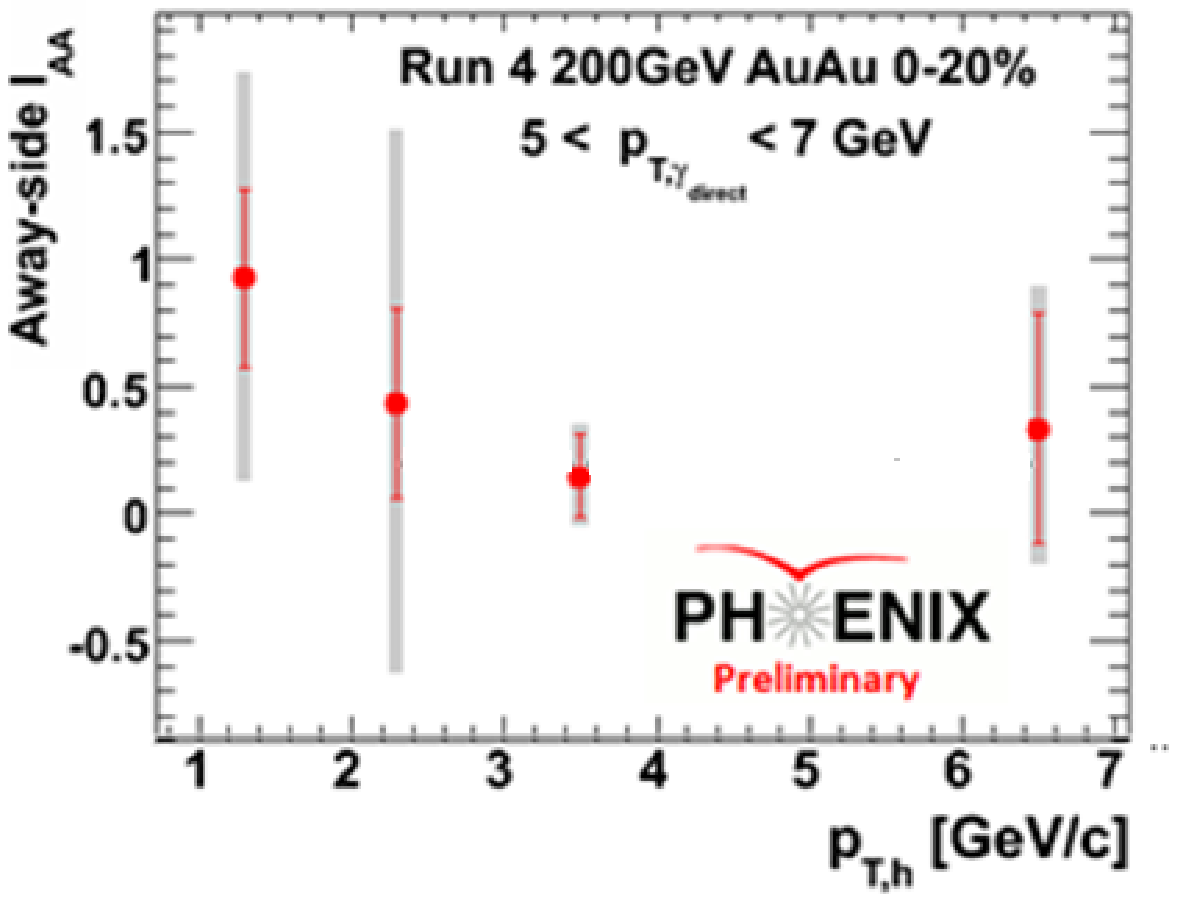}
\caption{Ratio \iaa\ (see text) for the lowest trigger direct photon \pt\ bin.}
\label{fig:iaa57}
\end{figure}

In previous conferences \cite{jiamin,matt} PHENIX has presented \auau\ results of the statistical subtraction method for direct photon-hadron correlation yields.  With the now statistically improved $p+p$ results (for this section, using the pure statistical subtraction method in \pp\ for comparison, combining Run6 and Run6 statistics) and an expanded \pt\ range for the trigger photons in $Au+Au$, we can now divide the per-trigger yields for \auau\ and \pp in many \pt\ bins making the ratio \iaa\ = $Y_{direct}^{A+A}/Y_{direct}^{p+p}$ for an awayside integration range of $2\pi/5$ radians around $\pi$.   An example is shown for the direct photon (trigger) \pt\ bin 5-7 GeV/c in figure \ref{fig:iaa57}. It is apparent that there are very large uncertainties, but most points have positive yield in \auau\ and a value of \iaa\ between 0 and 1 which indicates the basic expectation of suppression of the awayside jet.

We find that different sources of uncertainties dominate in different ranges of trigger photon and associated hadron hadron \pt.  As either \pt\ is increased the statistics in the raw correlation functions obviously decrease as the production probabilities follow steeply falling spectra.  However the combinatoric background from pairs of hard particles and soft underlying or otherwise uncorrelated particles is reduced dramatically as the \pt\ of the associated hadron is increased, having the effect of reducing the systematic uncertainty from the underlying event subtraction. Further, as the \pt\ of the trigger photon is increased there is a larger fraction of direct photon-h pairs in the inclusive photon-h pair sample due to the increasing direct photon signal relative to the suppressed decay photon background, and thus the higher trigger photon \pt\ bins also have a reduction in the uncertainty from the subtraction in equation \ref{eq:subform}.  Because of these effects we find that the total uncertainty remains more constant with increasing \pt\ than the loss in statistical precision that the falling production probability would normally manifest. For this reason, integrating over large \pt\ hadron or trigger \pt\ bins causes a loss of information since the steeply falling production probability causes uncertainties to be dominated by the source corresponding to the low end of the \pt\ bin.  Therefore in order to effectively give the higher \pt\ bins more weight, we take the plain average of all \ptgpth\ bins.   We call
this average the Mean Value $I_{AA}$ akin to performing the functional mean value:

\begin{equation}
Mean Value\;\;I_{AA} = \frac{1}{{\Delta p_T }}\int {I_{AA} (p_T^{assoc} )dp_T }
\end{equation}

Under the further assumption that \iaa\ remains constant with \pt\, this average indeed corresponds exactly to the functional mean value, and it is found that this assumption is satisfied to what is likely a sufficient degree (considering our large uncertainties which would dominate any such uncertainties in the assumption) both in measurements of di-hadron correlations \cite{phenix_ppg083} and \cite{stardijet} and in most theoretical predictions \cite{arleo,renk} at sufficiently high \pt\ (associated \pt\ $> \simeq$ GeV/c).

\begin{figure*}
\includegraphics[width=1.0\linewidth]{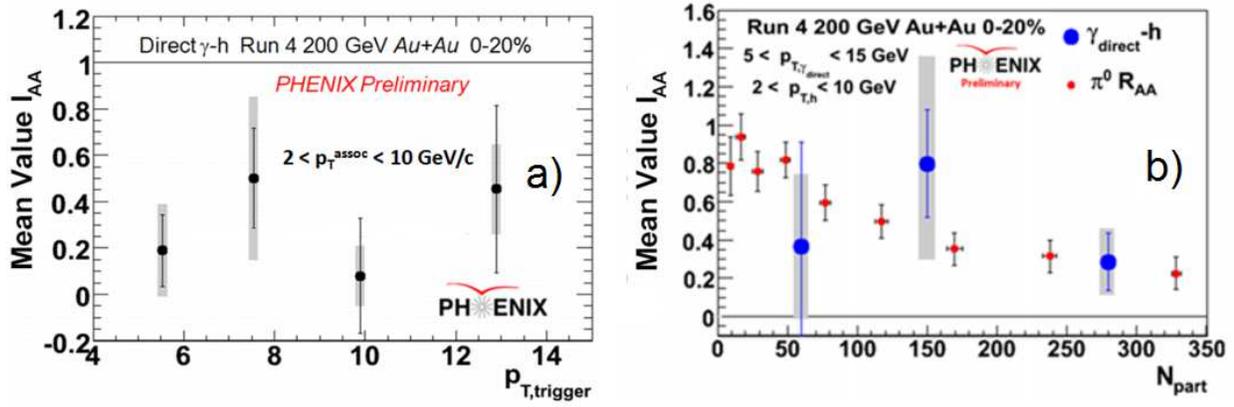}
\caption{a) Mean value \iaa\ for each trigger \pt\ bin b) over all \pt\ bins for 0-20\%, 20-40\%, and 40-92\% centrality bins.}
\label{fig:meanvalue_iaa}       
\end{figure*}

Results of the Mean Value \iaa\ are shown in figure \ref{fig:meanvalue_iaa}.   The transition of different uncertainty sources from systematic to statistical is even more apparent in figure \ref{fig:meanvalue_iaa} $a)$ for central events, and indeed the Mean Value \iaa\ seems to favor a constant value \emph{vs.}
trigger direct photon \pt\, although the uncertainties are still much too large to rule out with any significance rather large possible trigger photon \pt\ dependence scenarios.  Still, again taking the average of \emph{all} \iaa\ points together, shown on figure
\ref{fig:meanvalue_iaa} $b)$, now also for more peripheral bins, we find that the Mean Value \iaa\ for the central events is significantly positive at a two-sigma level and consistent with the single particle suppression level \raa.

This confirms the basic geometrical picture of energy loss at RHIC \cite{renkgeo},\cite{xnwanggeo} because vertices in $\gamma$-h correlations, and the trigger direct photon themselves, come from the entire collision volume.  Therefore the fraction of \gam-$h$ pairs that are observed without significant suppression, come from the same geometrical region that surviving single high \pt\ particles do. Dividing this quantity by the unsuppressed direct photon yield is analogous to dividing the surviving single particle production rate by the expectation from $p+p$ multiplied over the entire production weighted geometric volume of the initial distribution of all hard scattering production points, as is done in the construction of \raa\.  This means that the away-side suppression in the $\gamma-jet$ channel should simply reflect the same geometry as the single particle picture and, if geometry plays a dominant role, give the same suppression level.  Since this geometrical picture is believed to be a "surface bias" picture where the only di-jets initially produced near the surface it could be a confirmation of the surface bias picture, although further comparison to the di-hadron \iaa\, comparing $\gamma$-hadron and di-hadron correlations together in the same energy loss framework, are necessary to make this statement, in order to rule out possible effects in the di-hadron \iaa\ that are believed \emph{not} to be from geometry.

\begin{figure}[tb]
\centering
\includegraphics[width=0.5\textwidth]{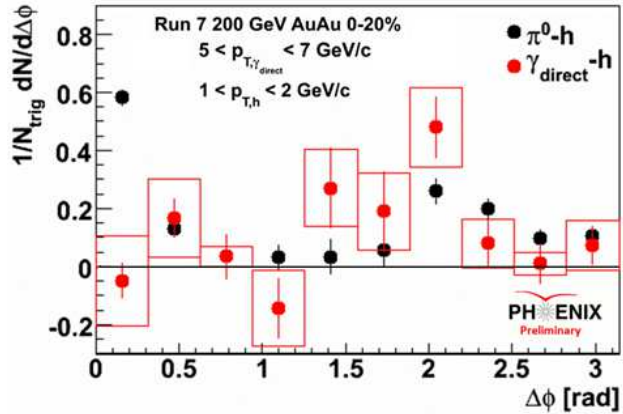}
\caption{$\Delta\phi$ distribution of direct photon-hadron correlations from Run7 data for 5-7 trigger direct photon bin.}
\label{fig:run7}
\end{figure}

Looking forward to the full release of our Run7 analysis, we also present a first look in figure \ref{fig:run7} at the lowest trigger \pt\ bin using $\simeq$ 3.0 billion events, which entices us to look across several \pt\ bins to look for consistent behavior that may be consistent to jet-medium displaced peak as seen already in di-hadron correlations, although at this point uncertainties are too large to make a definitive statement with just a single \pt\ bin combination.

\section{Conclusions}
\label{sec:conclusions}

Various 2-particle direct photon-hadron correlation yields in $p+p$ and $Au+Au$ collisions at \sqrtsNN= 200 GeV have been presented.  Under the assumption that the suppression is nearly \pt\ independent using a specific averaging scheme, we find a ratio of \auau\ to \pp\ per-trigger photon yields averaged over all available \pt\ bins, $I_{AA}$, consistent with the single particle suppression level \raa, which can be interpreted as a qualitative confirmation of the basic geometrical picture of jet suppression at RHIC.  To the extent that the prevailing geometric picture is believed to be that of surface bias where only jets ejected near the surface dominately contribute to yields, \cite{renkgeo}, \cite{xnwanggeo}, it may be said to confirm this picture, however it should be noted that any geometric scenario would yield \iaa\ $\simeq$ \raa.  Nonetheless future comparisons of \iaa\ from \gam-jet to that of di-jets, may indeed yield some insight into the details of possible surface bias models.

We have also presented the first event by event isolation cut 2-p correlation results at RHIC.  The application of the event by event photon isolation cuts in $p+p$ results our highest precision measurement yet, and allows for precision studies of the baseline fragmentation function $D(z)$, and well as a variable \pout\ which is proportional to the apparent intrinsic $k_T$, or non-zero transverse momentum of the original collision partons.  With a model dependent extraction method, the average $\sqrt{<k_T^2>}$ at this \sqrts\ in \pp\ is found to be approximately 2.5-3 GeV, consistent with analysis of di-hadron (di-jet) correlations \cite{ppg029}.  The possible implications of this and also the improved precision in the isolated yields warrant detailed comparison with the baseline perturbative QCD (pQCD) calculations used in the various models of jet energy loss.  Finally, we have presented a unique direct measurement of single prompt photons from jet fragmentation, of both angular information and the fraction of these photons in the entire photon sample (including decay photons) in the vicinity of the jet cone and for specific \pt\ cuts.

%
%

\end{document}